\documentclass[a4paper,11pt]{article}

\usepackage{jheppub}
\allowdisplaybreaks
\usepackage{graphicx}
\usepackage{epstopdf}

\newcommand{\be}{\begin{equation}}
\newcommand{\ee}{\end{equation}}
\newcommand{\bea}{\begin{eqnarray}}
\newcommand{\eea}{\end{eqnarray}}
\newcommand{\nn}{\nonumber}

\title{Leading order finite size effects with spins \\
for inspiralling compact binaries}

\author[a,b]{Michele Levi}
\author[c,d]{and Jan Steinhoff}

\affiliation[a]{Universit\'e Pierre et Marie Curie-Paris VI, CNRS-UMR 7095, 
Institut d'Astrophysique de Paris, \\ 98 bis Boulevard Arago, 75014 Paris, France} 
\affiliation[b]{Sorbonne Universit\'es, Institut Lagrange de Paris, \\ 
98 bis Boulevard Arago, 75014 Paris, France} 

\affiliation[c]{Max-Planck-Institute for Gravitational Physics 
(Albert-Einstein-Institute),\\ Am M{\"u}hlenberg 1, 14476 Potsdam-Golm, Germany}
\affiliation[d]{Centro Multidisciplinar de Astrofisica, Instituto Superior Tecnico, 
Universidade de Lisboa,\\ Avenida Rovisco Pais 1, 1049-001 Lisboa, Portugal}

\emailAdd{michele.levi@upmc.fr}
\emailAdd{jan.steinhoff@aei.mpg.de}

\abstract{The leading order finite size effects due to spin, 
namely that of the cubic and quartic in spin interactions, are 
derived for the first time for generic compact binaries via 
the effective field theory for gravitating spinning objects. These corrections 
enter at the third and a half and fourth post-Newtonian orders, respectively,
for rapidly rotating compact objects. Hence, we complete the leading order 
finite size effects with spin up to the fourth post-Newtonian accuracy. 
We arrive at this by augmenting the point particle effective action with new 
higher dimensional nonminimal coupling worldline operators, involving 
higher-order derivatives of the gravitational field, and introducing new 
Wilson coefficients, corresponding to constants, which describe the octupole 
and hexadecapole deformations of the object due to spin. 
These Wilson coefficients are fixed to unity in the black hole case.
The nonminimal coupling worldline operators enter the action 
with the electric and magnetic components of the Weyl tensor of even and odd parity, 
coupled to even and odd worldline spin tensors, respectively.
Moreover, the non relativistic gravitational field decomposition, 
which we employ, demonstrates a coupling hierarchy of the 
gravito-magnetic vector and the Newtonian scalar, to the odd 
and even in spin operators, respectively, which extends 
that of minimal coupling. This observation is useful 
for the construction of the Feynman diagrams, and provides 
an instructive analogy between the leading order spin-orbit 
and cubic in spin interactions, and between the leading order 
quadratic and quartic in spin interactions.}

\begin{document}

\maketitle

\flushbottom

\section{Introduction}

Second-generation ground-based interferometers, such as Advanced LIGO \cite{LIGO}, 
Advanced Virgo \cite{Virgo}, and KAGRA \cite{Kagra}, will start to operate in the 
next few years, making the anticipated direct detection of gravitational waves (GWs) 
a realistic prospect. Among the most promising sources in the accessible frequency band 
of such experiments are inspiralling binaries of compact objects, which can be treated 
analytically in terms of the post-Newtonian (PN) approximation of General Relativity 
\cite{Blanchet:2013haa}. It turns out that even relative high order corrections beyond 
Newtonian gravity, such as the fourth PN (4PN) order, are crucial to obtain a successful 
detection from such sources, and furthermore to gain information about the inner 
structure of the constituents of the binary \cite{Yagi:2013baa}. 
Moreover, such objects are expected to have large spins \cite{McClintock:2011zq}, 
thus PN corrections involving spins should be completed to similar high orders as 
in the non spinning case, which was recently completed to 4PN order \cite{Damour:2014jta}.

In particular, finite size effects involving spins should also be taken 
into account in order to obtain the required 4PN accuracy. 
The leading order (LO) finite size spin effects at the quadrupole level, 
i.e.~of the LO spin-squared interaction, were first derived for black holes 
in \cite{D'Eath:1975vw, Thorne:1984mz}. Generic quadrupoles, required to describe 
neutron stars of different masses, were included 
already in \cite{Barker:1975ae}, and the proportionality of the quadrupole 
to spin-squared was introduced in \cite{Poisson:1997ha}. 
The LO spin-squared interaction enters already at the 2PN order 
for rapidly rotating generic compact objects. 
Yet, the next-to-leading order (NLO) spin-squared interaction at 3PN order 
was treated much later in the following series of works 
\cite{Porto:2008jj, Steinhoff:2008ji, Hergt:2008jn, Hergt:2010pa}. 
Finally, the LO cubic and quartic in spin interaction Hamiltonians 
for black hole binaries were computed in parts in 
\cite{Hergt:2007ha,Hergt:2008jn}. These corrections enter formally at the 2PN 
order, and for rapidly rotating compact objects at the 3.5PN and 4PN orders, 
respectively. However, these results were found to be incomplete in the 
test particle limit for the quartic in spin sector \cite{Steinhoff:2012rw}. 

In this work we derive for the first time the LO cubic and 
quartic in spin interaction potentials for generic compact binaries 
via the Effective Field Theory (EFT) for gravitating spinning objects 
\cite{Levi:2015msa}. Hence, we complete 
the LO finite size effects with spin up to the 4PN accuracy.
The novel self-contained EFT approach for the binary inspiral problem was 
introduced in \cite{Goldberger:2004jt,Goldberger:2007hy}, and 
an extension to spinning objects was first approached in \cite{Porto:2005ac}. 
The EFT approach provides a systematic methodology to construct the 
action to arbitrary high accuracy in terms of local operators ordered 
by their relevance and their Wilson coefficients, which is invaluable 
for the obtainment of finite size effects. Moreover, the EFT approach 
also applies the efficient standard tools of Quantum Field Theory, 
such as Feynman diagrams. Indeed, we arrive at our results by augmenting 
the point particle action with new higher dimensional nonminimal 
coupling worldline operators, involving higher-order derivatives of the 
field, and introducing new Wilson coefficients, namely here spin-induced polarization coefficients, corresponding to constants, 
which describe the octupole and hexadecapole deformations of the object 
due to spin \cite{Levi:2015msa}. These Wilson coefficients are fixed to 
unity in the black hole case. The nonminimal coupling worldline operators 
enter the action with the electric and magnetic components of the Weyl tensor
of even and odd parity, coupled to even and odd worldline spin tensors, respectively.

Moreover, another advantageous practice in the EFT approach 
is the use of the non relativistic gravitational (NRG) fields, which were 
introduced in \cite{Kol:2007bc}. The NRG decomposition of spacetime is 
essentially a reduction over the time dimension, and therefore it is the 
sensible decomposition for the PN limit \cite{Kol:2010ze}.
Indeed, the NRG field decomposition, which we employ, demonstrates 
a coupling hierarchy of the gravito-magnetic vector and the 
Newtonian scalar to the odd and even in spin operators, respectively, 
which extends that of minimal coupling, as was illustrated for the spin 
interactions in \cite{Levi:2008nh,Levi:2010zu,Levi:2011eq}. 
This observation is very useful for the construction of the Feynman diagrams, 
and provides an instructive analogy between the LO spin-orbit and cubic in 
spin interactions, and between the LO quadratic and quartic in spin 
interactions.
  
The outline of the paper is as follows. In section \ref{fsseft} we review and 
present the new ingredients in the EFT formulation for finite size effects 
with spins to the order required in this work \cite{Levi:2015msa}. In section 
\ref{locsi} we derive the complete LO cubic in spin interaction potential for 
generic compact objects through the evaluation of the relevant Feynman diagrams, 
and we compare to the ADM Hamiltonian result for a black hole 
binary. In section \ref{loqsi} we similarly derive the complete LO quartic in 
spin interaction potential for generic compact objects, and we correct the 
corresponding ADM Hamiltonian result for the black hole binary case. Finally, 
in section \ref{theendmyfriend} we summarize our main conclusions.

Throughout this paper we use $c\equiv1$, 
$\eta_{\mu\nu}\equiv \text{Diag}[1,-1,-1,-1]$,
and the convention for the Riemann tensor is 
$R^\mu{}_{\nu\alpha\beta}\equiv\partial_\alpha\Gamma^\mu_{\nu\beta}
-\partial_\beta\Gamma^\mu_{\nu\alpha}
+\Gamma^\mu_{\lambda\alpha}\Gamma^\lambda_{\nu\beta}
-\Gamma^\mu_{\lambda\beta}\Gamma^\lambda_{\nu\alpha}$. 
Greek letters denote indices in the global coordinate frame. 
Spatial tensor indices are denoted with lowercase Latin letters 
from the middle of the alphabet, whereas uppercase ones denote particle labels.
The scalar triple product appears here with no brackets, 
i.e.~$\vec{a}\times\vec{b}\cdot\vec{c}\equiv(\vec{a}\times\vec{b})\cdot\vec{c}$.

\section{Spin-induced finite size effects via the effective field theory for spin} 
\label{fsseft}

In this section we present the effective action, that removes the scale of 
the compact objects in the EFT approach, and augment the point particle action, 
as required in order to take into account finite size effects with spins. 
For the construction of the spin-induced nonminimal couplings we 
follow the EFT for spin in \cite{Levi:2015msa}, where an equivalent approach for generic tidal 
interactions is found in \cite{Bini:2012gu}.
For the Feynman rules we employ here 
the NRG fields \cite{Kol:2007bc,Kol:2010ze}, which continue to play a 
central role in the construction of the Feynman diagrams also in spin 
interactions, as will be seen in the next sections. Thus, their central role in 
spin interactions extends beyond the minimal coupling case, which was 
illustrated in \cite{Levi:2008nh,Levi:2010zu,Levi:2011eq}.  

We recall that the effective action, describing the binary system, is given by
\be \label{totact}
S=S_{\text{g}}+\sum_{I=1}^{2}S_{\text{(I)pp}},
\ee
where $S_{\text{g}}$ is the pure gravitational action, and $S_{\text{(I)pp}}$ is 
the worldline point particle action for each of the two particles in the binary. 
The gravitational action is the usual Einstein-Hilbert action plus 
a gauge-fixing term, which we choose as the fully harmonic gauge, 
such that we have
\be
S_{\text{g}}=S_{\text{EH}} + S_{\text{GF}} 
= -\frac{1}{16\pi G} \int d^4x \sqrt{g} \,R + \frac{1}{32\pi G} \int d^4x\sqrt{g}
\,g_{\mu\nu}\Gamma^\mu\Gamma^\nu, 
\ee
where $\Gamma^\mu\equiv\Gamma^\mu_{\rho\sigma}g^{\rho\sigma}$.

In terms of the NRG fields $\phi$, $A_i$, $\gamma_{ij}\equiv\delta_{ij}+\sigma_{ij}$, 
the metric reads
\begin{align} \label{eq:gkk}
g_{\mu\nu}&=
\left(\begin{array}{cc} 
e^{2\phi}      & \quad -e^{2\phi} A_j \\
-e^{2\phi} A_i & \quad -e^{-2\phi}\gamma_{ij}+e^{2\phi} A_i A_j
\end{array}\right)&\simeq
\left(\begin{array}{cc} 
1+2\phi & \quad -A_j \\
-A_i    & \quad -\delta_{ij}+2\phi\delta_{ij}
\end{array}\right), 
\end{align}
where we have written the approximation in the weak field limit up to linear order 
in the fields as required in this work. 

The NRG scalar and vector field propagators in the harmonic gauge are given by 
\begin{align}
\label{eq:prphi} \parbox{18mm}{\includegraphics[scale=0.6]{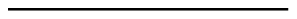}}
 = & \langle{~\phi(x_1)}~~{\phi(x_2)~}\rangle = ~~~~4\pi G~~~ \int
 \frac{d^d\vec{k}}{\left(2\pi\right)^d}\frac{e^{i\vec{k}\cdot
 \left(\vec{x}_1 - \vec{x}_2\right)}}{k^2}~\delta(t_1-t_2),\\ 
\label{eq:prA} \parbox{18mm}{\includegraphics[scale=0.6]{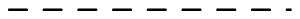}}
 = & \langle{A_i(x_1)}~{A_j(x_2)}\rangle = -16\pi G~\delta_{ij} \int
 \frac{d^d\vec{k}}{\left(2\pi\right)^d}
 \frac{e^{i\vec{k}\cdot\left(\vec{x}_1 -\vec{x}_2\right)}}{k^2}~\delta(t_1-t_2). 
\end{align}

Next, we recall that the minimal coupling part of the point particle action 
of each of the particles with spins 
\cite{Hanson:1974qy, Bailey:1975fe,Porto:2005ac,Levi:2010zu} is given by  
\begin{align} \label{mcact}
S_{\text{pp}}=&\int 
d\lambda\left[-m \sqrt{u^2}-\frac{1}{2} S_{\mu\nu}\Omega^{\mu\nu}\right],
\end{align}
where $\lambda$ is the affine parameter, $u^{\mu}\equiv dx^\mu/d\lambda$ is the 4-velocity,
and $\Omega^{\mu\nu}$, $S_{\mu\nu}$ are the angular velocity and spin tensors of the 
particle, respectively \cite{Levi:2015msa}.
Considering this point particle action in eq.~\eqref{totact}, the LO 
monopole-monopole interaction, namely the Newtonian interaction, 
which involves no spin, and the LO dipole-monopole interaction, 
namely the linear in spin LO spin-orbit interaction, are derived. 
These are obtained using the following Feynman rules we present, 
which are also those required to the order that we are considering in this 
work. Here we follow the same gauge choices and conventions as in \cite{Levi:2015msa}, 
see section 5.3 there.

For the one-graviton couplings to the worldline mass, we have
\begin{align}
\label{eq:mphi} \parbox{12mm}{\includegraphics[scale=0.6]{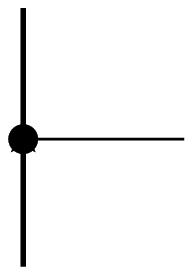}}
  = & - m \int dt~\phi, \\ 
\label{eq:mA} \parbox{12mm}{\includegraphics[scale=0.6]{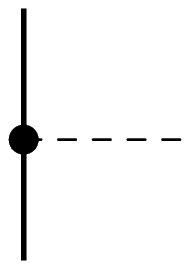}}
  = & ~m \int dt~A_iv^i, 
\end{align}
where the heavy solid lines represent the worldlines, and 
the spherical black blobs represent the masses on the worldline. 

The Feynman rules for the one-graviton couplings to the worldline spin are: 
\begin{align}
\label{eq:sA}  \parbox{12mm}{\includegraphics[scale=0.6]{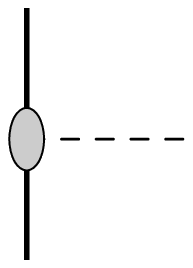}}
 = & \frac{1}{2} \int dt \,\,\left[\epsilon_{ijk}S_{k}\partial_iA_j \right], \\ 
\label{eq:sphi}   \parbox{12mm}{\includegraphics[scale=0.6]{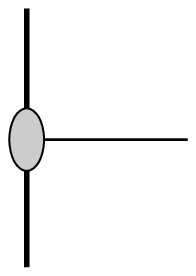}}
 = & 2 \int dt \,\, \left[\epsilon_{ijk}S_{k} v^i \partial_j\phi  \right],   
\end{align}
where $\epsilon_{ijk}$ is the 3-dimensional Levi-Civita symbol, such that 
$S_{ij}=\epsilon_{ijk}S_{k}$, and the oval gray blobs represent the spins on the 
worldline. Note that the Feynman rules here contain only spatial components of the 
spin tensor, since the rotational gauge fixing is applied at the level of the 
point particle action, and thus the unphysical degrees of freedom of spin are 
already eliminated. Our rotational variables are ultimately fixed to the 
canonical gauge \cite{Levi:2015msa}. 

In order to take into account finite size effects with spins the point 
particle action should be extended beyond minimal coupling, see section 4 of 
\cite{Levi:2015msa} for the complete treatment. 
Higher dimensional operators are introduced into the action, constructed 
with the Riemann tensor, which for a vacuum solution field is equivalent to the Weyl tensor, 
using its even and odd parity components.
$E_{\mu\nu}$ is the electric component of the Weyl tensor, namely 
\be
E_{\mu\nu}\equiv R_{\mu\alpha\nu\beta} u^{\alpha}u^{\beta},
\ee
and $B_{\mu\nu}$ is the magnetic component of the Weyl tensor, that is 
\be
B_{\mu\nu}\equiv\frac{1}{2}\epsilon_{\alpha\beta\gamma\mu} 
R^{\alpha\beta}_{\quad\delta\nu} u^{\gamma}u^{\delta},
\ee 
where $\epsilon_{\alpha\beta\gamma\mu}$ is the Levi-Civita tensor.
The LO PN finite size effects are due to spin-induced multipoles, for which 
only linear in curvature operators should be considered. 

These higher order operators are constructed according to the symmetries of 
the effective action \cite{Levi:2015msa}. The crucial symmetries to consider 
for the construction of nonminimal couplings are parity invariance, and SO(3) 
invariance of the body-fixed triad, where we start with a timelike basis vector, 
satisfying $e_{[0]}{}^{\mu}=u^\mu/\sqrt{u^2}$, which amounts to a covariant 
gauge condition for the worldline tetrad. 
We also recall here that permanent multipole moments, beside mass and spin, are assumed to vanish, and that parity violation can be neglected for macroscopic compact objects in general relativity.
The building blocks of these higher order
operators are then spin-induced multipoles and curvature components with covariant 
derivatives, which are considered in the body-fixed frame, where they form 
irreducible representations of SO(3). Due to parity invariance the tensors, 
which contain the even-parity electric and odd-parity magnetic curvature 
components, should be contracted with an even and odd number of the 
spin vector, respectively, of an equal tensor rank, where the spin vector 
$S^{\mu}$ is defined here by
\be
S^{\mu}\equiv\frac{1}{2}\epsilon_{\alpha\beta\gamma\mu}S^{\alpha\beta}
\frac{u^{\gamma}}{\sqrt{u^2}}.
\ee
Finally, these nonminimal couplings carry Wilson coefficients, which encode the 
internal structure of the object. 

For the LO spin-squared finite size effects the nonminimal coupling operator, 
which is added to the action \cite{Levi:2015msa}, is given by
\begin{align} \label{es2act}
L_{\text{ES}^2}=& 
-\frac{C_{\text{ES}^2}}{2m}\frac{E_{\mu\nu}}{\sqrt{u^2}}S^{\mu} S^{\nu}.
\end{align}
$C_{\text{ES}^2}$ is the Wilson coefficient \cite{Porto:2008jj}, corresponding 
to the quadrupole deformation of the object due to spin, which was introduced in 
\cite{Poisson:1997ha}, where the proportionality factor was called $a$. The 
spin-squared operator presented in eq.~\eqref{es2act} is equivalent to that in 
\cite{Porto:2008jj}, as the building blocks of the spin-induced nonminimal 
couplings, considered in the body-fixed frame, are traceless and orthogonal to the 
4-velocity \cite{Levi:2015msa}.

Considering this addition to the point particle action in eq.~\eqref{mcact} 
and eq.~\eqref{totact}, the LO quadrupole-monopole interaction, 
that is the LO spin-squared interaction, is derived, using the following Feynman rules we 
present, required to the order that we are considering in this work.
The Feynman rules for the one-graviton couplings to the worldline spin-squared 
are given by
\begin{align}
\label{eq:sqphi}   \parbox{12mm}{\includegraphics[scale=0.6]{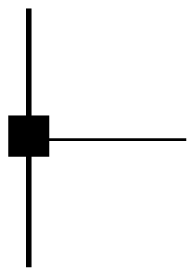}}
 =& \int dt \,\,\left[\frac{C_{\text{ES}^2}}{2m}\left(S^{i}S^{j}\partial_i\partial_j\phi-S^2\partial_i\partial_i\phi\right)\right] ,\\
\label{eq:sqA}  \parbox{12mm}{\includegraphics[scale=0.6]{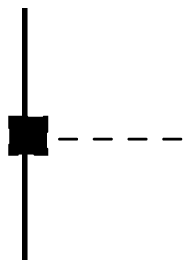}}
 =& \int dt \,\, \left[-\frac{C_{\text{ES}^2}}{2m}\left(S^{i}S^{j}\left(\partial_i\partial_j A_lv^l
 -\partial_i\partial_l A_jv^l-\partial_i\partial_tA_j\right)\right.\right.\nn\\
 & \qquad\qquad\qquad\left.\left.-S^2\left(\partial_i\partial_i A_kv^k
 -\partial_i\partial_k A_iv^k-\partial_i\partial_tA_i\right)\right)\right],
\end{align}
where the square black boxes represent the $ES^2$ spin-squared operators on the 
worldlines. Note that in spite of naive power counting only the first terms in eqs.~\eqref{eq:sqphi}, \eqref{eq:sqA}, actually contribute here at LO. Using the leading coupling of the spin-squared to the Newtonian 
scalar in eq.~\eqref{eq:sqphi}, contracted with the corresponding leading mass 
coupling in eq.~\eqref{eq:mphi}, we obtain the single Feynman diagram, which makes up 
the well-known LO spin-squared interaction \cite{Barker:1975ae,Poisson:1997ha}, shown in 
figure 1. 
\begin{figure}[t]
\begin{center}
\includegraphics[scale=0.75]{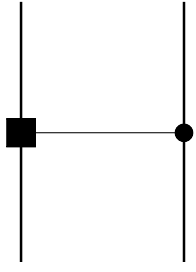}
\caption{LO spin-squared interaction Feynman diagram. 
This diagram should be included together with its mirror image.
This is a quadrupole-monopole interaction. 
Note the analogy with the Newtonian interaction.} 
\end{center}
\label{fig:S2LO}
\end{figure}
The value of the Feynman diagram for the LO spin-squared interaction is given by
\begin{align}
\text{Fig.~1}=&\frac{1}{2}C_{1\text{(ES$^2$)}}\frac{Gm_2}{m_1r^3}
\left[S_1^2-3\left(\vec{S}_1\cdot\vec{n}\right)^2\right],
\end{align}
where $\vec{r}\equiv\vec{x}_1-\vec{x}_2$, $r\equiv\sqrt{{\vec{r}}^{2}}$, 
$\vec{n}\equiv\vec{r}/r$, and the LO spin-squared potential 
just equals $V_{\text{S$_1^2$}}^{\text{LO}}=-\text{Fig.~1}$. 
Notice that this is a purely Newtonian effect, and that the worldline 
spin-squared acts just like a generic mass quadrupole. 

In this work we want to complete the LO octupole and hexadecapole levels in 
the spins, since these contribute up to the 4PN accuracy for rapidly rotating 
compact objects. For that, the point particle action should be extended to LO 
cubic and quartic order in the spin \cite{Levi:2015msa}. 

The cubic in spin operator, that should be added here \cite{Levi:2015msa}, is given by
\begin{align}
L_{\text{BS}^3}=&-\frac{C_{\text{BS}^3}}{6m^2}\frac{D_{\lambda} B_{\mu\nu}}{\sqrt{u^2}}
S^{\mu} S^{\nu}S^{\lambda},
\end{align}
where $D_{\lambda}$ denotes the covariant derivative. We have introduced here 
$C_{\text{BS}^3}$, which is the Wilson coefficient, or constant describing the 
octupole deformation due to spin.

The Feynman rules for the one-graviton couplings to the 
worldline cubic spin are then given by
\begin{align}
\label{eq:s3A}  \parbox{12mm}{\includegraphics[scale=0.6]{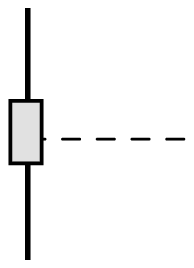}}
 =& \int dt \,\, \left[-\frac{C_{\text{BS}^3}}{12m^2}
 S^{i}S^{j}\epsilon_{klm}S^{m}\partial_i\partial_j\partial_k A_l\right],\\ 
\label{eq:s3phi}   \parbox{12mm}{\includegraphics[scale=0.6]{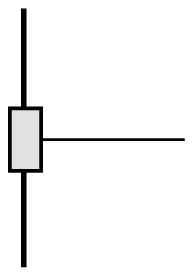}}
 =& \int dt \,\,\left[\frac{C_{\text{BS}^3}}{3m^2}
 S^{i}S^{j}\epsilon_{klm}S^{m}\partial_i\partial_j\partial_k\phi\,v^l\right],
\end{align}
where the rectangular gray boxes represent the $BS^3$ 
cubic spin operators on the worldlines.
Here we have only included terms, which contribute at this order.
Note that here it is the gravito-magnetic vector, which is the leading 
one in the hierarchy of coupling in the magnetic component 
of the Weyl tensor in the worldline cubic spin operator. 

The quartic in spin operator, that should be added here \cite{Levi:2015msa}, is given by
\be
L_{\text{ES}^4}=\frac{C_{\text{ES}^4}}{24m^3} 
\frac{D_{\kappa}D_{\lambda}E_{\mu\nu}}{\sqrt{u^2}} 
S^{\mu} S^{\nu} S^{\lambda} S^{\kappa}.
\ee
We have introduced here $C_{\text{ES}^4}$, which is the Wilson coefficient, 
or constant describing the hexadecapole deformation due to spin.

Then, the Feynman rule for the one-graviton coupling to the 
worldline quartic spin is given by
\begin{align}
\label{eq:s4phi}   \parbox{12mm}{\includegraphics[scale=0.6]{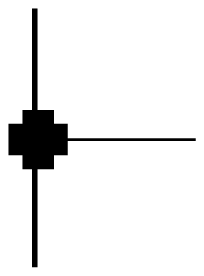}}
 =& \int dt \,\,\left[-\frac{C_{\text{ES}^4}}{24m^3}S^{i}S^{j}S^{k}S^{l}
 \partial_i\partial_j\partial_k\partial_l\phi\right],
\end{align} 
where the crossed black box represents the $ES^4$ quartic spin 
operator on the worldline. 
Here again we have omitted terms, which actually only contribute beyond LO.
Note that here it is again the Newtonian scalar, which is the 
leading one in the hierarchy of coupling in the electric component 
of the Weyl tensor in the worldline quartic spin operator.

\section{Leading order cubic in spin interaction}\label{locsi}

The LO cubic and quartic in spin interaction Hamiltonians for the 
black hole binary case were approached in parts in \cite{Hergt:2007ha,Hergt:2008jn}. 
These corrections enter formally at the 2PN order, and at the 
3.5PN and 4PN orders, respectively, for rapidly rotating compact 
objects. In this section and the next we derive these interaction 
potentials for any generic compact binary from the EFT for spin \cite{Levi:2015msa}, 
where we construct these complete interactions in a direct and instructive manner. 

\subsection{Feynman diagrams}

The cubic in spin interaction contains two kinds of interaction: 
a quadrupole-dipole interaction, and an octupole-monopole one.
Each of these two interactions is analogous to the LO spin-orbit 
interaction, which is a dipole-monopole interaction.
The correspondence is between even and odd parity multipole moments of 
the spin, such that the quadrupole and octupole moments correspond to 
the monopole (mass) and dipole (spin), respectively.
We recall from figure 1 in section IV of \cite{Levi:2010zu}, 
that the LO spin-orbit interaction contains two contributing Feynman diagrams, 
mediated by one-graviton exchanges of the gravito-magnetic vector 
and the Newtonian scalar of the NRG fields. Therefore, we expect 
to have here four contributing Feynman diagrams, two for each of 
the two kinds of interaction, that make up the cubic in spin interaction. 

Indeed, the four contributing Feynman diagrams can be seen here 
in figure 2, where on the left diagrams (a) and (b), we have the 
quadrupole-dipole interaction, and on the right diagrams (c) and (d), 
we have the octupole-monopole interaction. These diagrams are obtained 
by the following contractions: in figure 2(a) the LO worldline 
spin coupling to the gravito-magnetic vector from eq.~\eqref{eq:sA} 
is contracted with the corresponding quadrupole coupling in eq.~\eqref{eq:sqA}; 
in figure 2(b) we contract the LO worldline spin quadrupole coupling 
to the Newtonian scalar from eq.~\eqref{eq:sqphi} with the corresponding 
spin coupling in eq.~\eqref{eq:sphi}; in figure 2(c) the LO worldline spin 
octupole coupling to the gravito-magnetic vector from eq.~\eqref{eq:s3A} 
is contracted with the corresponding mass coupling in eq.~\eqref{eq:mA}; 
finally, in figure 2(d) the LO worldline mass coupling to the Newtonian 
scalar from eq.~\eqref{eq:mphi} is contracted with the corresponding spin 
octupole coupling in eq.~\eqref{eq:s3phi}. 
 
\begin{figure}[t]
\begin{center}
\includegraphics[scale=0.75]{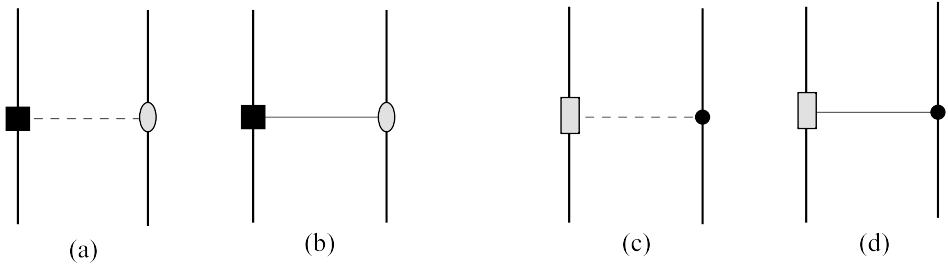}
\caption{LO cubic in spin interaction Feynman diagrams. These 
diagrams should be included together with their mirror images. 
On the left pair we have the quadrupole-dipole interaction, 
and on the right pair we have the octupole-monopole one. 
Note the analogy of each pair with the LO spin-orbit 
interaction in figure 1 of \cite{Levi:2010zu}.} 
\end{center}
\label{fig:S3LO}
\end{figure}

Hence, the values of the Feynman diagrams of the LO cubic in 
spin interaction are given by the following:
\begin{align}
\text{Fig.~2(a)}=& -3\frac{C_{1\text{(ES$^2$)}}G}{m_1r^4}
\left[S_1^2\,\vec{S}_2\cdot\vec{n}\times\vec{v}_1
+2\vec{S}_1\cdot\vec{n}\,\vec{S}_2\cdot\vec{S}_1\times\vec{v}_1
-5\left(\vec{S}_1\cdot\vec{n}\right)^2\vec{S}_2\cdot\vec{n}\times\vec{v}_1\right]\nn\\
&-3\frac{C_{1\text{(ES$^2$)}}G}{m_1r^3}\left[\dot{\vec{S}}_1\cdot\vec{n}\, 
\vec{n}\cdot\vec{S}_1\times\vec{S}_2+\vec{S}_1\cdot\vec{n}\, 
\vec{n}\cdot\dot{\vec{S}}_1\times\vec{S}_2\right],\label{diag2a}\\
\text{Fig.~2(b)}=& \,3\frac{C_{1\text{(ES$^2$)}}G}{m_1r^4}
\left[S_1^2\,\vec{S}_2\cdot\vec{n}\times\vec{v}_2
+2\vec{S}_1\cdot\vec{n}\,\vec{S}_2\cdot\vec{S}_1\times\vec{v}_2
-5\left(\vec{S}_1\cdot\vec{n}\right)^2\vec{S}_2\cdot\vec{n}\times\vec{v}_2\right],\\
\text{Fig.~2(c)}=& \,C_{1\text{(BS$^3$)}}\frac{Gm_2}{m_1^2r^4}
\vec{S}_1\cdot\vec{n}\times\vec{v}_2\left[S_1^2
-5\left(\vec{S}_1\cdot\vec{n}\right)^2\right],\\
\text{Fig.~2(d)}=& -C_{1\text{(BS$^3$)}}\frac{Gm_2}{m_1^2r^4}
\vec{S}_1\cdot\vec{n}\times\vec{v}_1\left[S_1^2
-5\left(\vec{S}_1\cdot\vec{n}\right)^2\right].
\end{align}
The evaluation of the diagrams here is straightforward. 
Yet, note that the value of diagram 2(b) depends on the choice of the spin gauge. 
Following \cite{Levi:2015msa}, after the covariant gauge is first inserted, and 
spin gauge freedom is incorporated in the action, an extra term from minimal 
coupling is added \cite{Levi:2015msa}, and is expected to supplement the potential here. 

\subsection{Effective potential and Hamiltonian}

As we noted in the end of the previous section, we recall 
that we have an addition from the extra minimal coupling term of 
\cite{Levi:2015msa}, coming from the incorporation 
of spin gauge freedom in the rotational minimal coupling term, 
which enters already at the LO spin-orbit sector. This addition 
contributes here only the kinematic piece, similar to that noted in eq.~(72) 
of \cite{Levi:2010zu}, given by  
\be
L_{\text{SSS(extra)}}^{\text{LO}}=
\frac{1}{2}\vec{S}_1\cdot\vec{v}_1\times\vec{a}_1 + 1\leftrightarrow2,
\ee 
and is acceleration dependent. We proceed then to eliminate 
the acceleration in this extra piece by making a redefinition of the position 
variables $\Delta\vec{x}_{I}$, as explained in \cite{Levi:2015msa}, according to
\begin{equation} \label{positionshift}
\vec{x}_1 \rightarrow \vec{x}_1 + \frac{1}{2 m_1}\vec{S}_1\times\vec{v}_1,
\end{equation}
and similarly for particle 2 with $1\leftrightarrow2$.
As explained in \cite{Damour:1990jh,Levi:2014sba} the linear shift in the position 
variables, corresponds to a substitution of the relevant EOM. Here these arise from 
the LO quadratic in spin sectors, and are given by
\begin{align}
m_1\vec{a}_{1\text{(SS)}}=&-\frac{3}{2}C_{1\text{(ES$^2$)}}
\frac{Gm_2}{m_1r^4}\left[\left(S_1^2-5\left(\vec{S}_1\cdot\vec{n}\right)^2\right)\vec{n}
+2\vec{S}_1\cdot\vec{n}\vec{S}_1\right]+ 1\leftrightarrow2\nn\\
& -\frac{3G}{r^4}\left[\left(\vec{S}_1\cdot\vec{S}_2
-5\vec{S}_1\cdot\vec{n}\vec{S}_2\cdot\vec{n}\right)\vec{n}
+\vec{S}_1\cdot\vec{n}\vec{S}_2+\vec{S}_2\cdot\vec{n}\vec{S}_1\right].
\end{align}
We obtain then the following addition: 
\begin{align} \label{extra}
L_{\text{SSS(extra)}}^{\text{LO}}=& \frac{3}{4}C_{1\text{(ES$^2$)}}\frac{Gm_2}{m_1^2r^4}
\left[\left(\vec{S}_1\cdot\vec{n}\times\vec{v}_1
-\frac{m_1}{m_2}\vec{S}_2\cdot\vec{n}\times\vec{v}_2\right)
\left(S_1^2-5\left(\vec{S}_1\cdot\vec{n}\right)^2\right)\right.\nn\\
&\left.-2\frac{m_1}{m_2}\vec{S}_1\cdot\vec{n}\vec{S}_2\cdot\vec{S}_1\times\vec{v}_2\right]
+\frac{3}{2}\frac{G}{m_1r^4}\left[\left(\vec{S}_1\cdot\vec{S}_2
-5\vec{S}_1\cdot\vec{n}\vec{S}_2\cdot\vec{n}\right)
\vec{S}_1\cdot\vec{n}\times\vec{v}_1\right.\nn\\
&\left.-\vec{S}_1\cdot\vec{n}\vec{S}_2\cdot\vec{S}_1\times\vec{v}_1\right]+ 1\leftrightarrow2.
\end{align}
Moreover, we note that the value of diagram 2(a) in eq.~\eqref{diag2a} contains higher order 
time derivatives of spin, of which a generic rigorous treatment was shown in 
\cite{Levi:2014sba}. According to this treatment a redefinition of the spin is required to 
remove its higher order time derivative, which amounts here to the insertion of the relevant EOM 
of the spin. Here these are the LO Newtonian EOM of the spin, given by
\be \label{seom}
\dot{S}^i=0,
\ee  
hence this insertion removes the terms with time derivative of the spin from eq.~\eqref{diag2a}.

Summing all diagrams in figure 2, with the extra addition in eq.~\eqref{extra}, and the 
insertion of eq.~\eqref{seom}, we get the following effective potential 
for the complete LO cubic in spin interaction:
\begin{align} \label{vsc}
V_{\text{SSS}}^{\text{LO}}=&C_{1\text{(BS$^3$)}}\frac{Gm_2}{m_1^2r^4}
\left(\vec{S}_1\cdot\vec{n}\times\vec{v}_1-\vec{S}_1\cdot\vec{n}\times\vec{v}_2\right)
\left(S_1^2-5\left(\vec{S}_1\cdot\vec{n}\right)^2\right)\nn\\
&+3\frac{C_{1\text{(ES$^2$)}}G}{m_1r^4}\left[\left(S_1^2\,\vec{S}_2\cdot\vec{n}\times\vec{v}_1
+2\vec{S}_1\cdot\vec{n}\,\vec{S}_2\cdot\vec{S}_1\times\vec{v}_1
-5\left(\vec{S}_1\cdot\vec{n}\right)^2\vec{S}_2\cdot\vec{n}\times\vec{v}_1\right)\right.\nn\\
&\qquad\qquad\quad-\left(S_1^2\,\vec{S}_2\cdot\vec{n}\times\vec{v}_2
+2\vec{S}_1\cdot\vec{n}\,\vec{S}_2\cdot\vec{S}_1\times\vec{v}_2
-5\left(\vec{S}_1\cdot\vec{n}\right)^2\vec{S}_2\cdot\vec{n}\times\vec{v}_2\right)\nn\\
&\left.-\frac{1}{4}\left[\left(\frac{m_2}{m_1}\vec{S}_1\cdot\vec{n}\times\vec{v}_1
-\vec{S}_2\cdot\vec{n}\times\vec{v}_2\right)
\left(S_1^2-5\left(\vec{S}_1\cdot\vec{n}\right)^2\right)
-2\vec{S}_1\cdot\vec{n}\vec{S}_2\cdot\vec{S}_1\times\vec{v}_2\right]\right]\nn\\
&-\frac{3}{2}\frac{G}{m_1r^4}\left[\left(\vec{S}_1\cdot\vec{S}_2
-5\vec{S}_1\cdot\vec{n}\vec{S}_2\cdot\vec{n}\right)\vec{S}_1\cdot\vec{n}\times\vec{v}_1
-\vec{S}_1\cdot\vec{n}\vec{S}_2\cdot\vec{S}_1\times\vec{v}_1\right]\nn\\
&+ 1\longleftrightarrow2.
\end{align}
The new Wilson coefficient $C_{\text{BS$^3$}}$ should 
be fixed in the black hole case. It is convenient to normalize
it to unity for black holes.
The binding energy can be used for a gauge invariant matching of all 
the Wilson coefficients encountered in this work.
A comparison of the gauge invariant binding energy, derived 
from our potential, with the one for a test-particle in the 
Kerr geometry \cite{Steinhoff:2012rw}, leads to $C_{\text{BS$^3$}} = 1$ 
for black holes. At the same time, this provides a check 
of our result against the small mass ratio case.
It should be stressed that this matching procedure
is based on the gauge invariant binding energy, and does not rely on a specific gauge dependent form of the metric.
We also note here the Geroch-Hansen mass multipoles $M_l$, and flux multipoles $S_l$, for black holes, given by \cite{Hansen:1974zz, Ryan:1995wh}
\begin{equation} \label{GHansen}
M_l + iS_l = M (ia)^l.
\end{equation}
Yet, the Geroch-Hansen multipoles must be related to the Wilson coefficients
by a matching calculation.

We would like to compare our effective potential to 
the ADM Hamiltonian results for a black hole binary (BHB)
derived in parts in \cite{Hergt:2007ha,Hergt:2008jn}. 
Collecting the pieces from eq.~(144) in \cite{Hergt:2007ha}, 
and eqs.~(7.1), (7.2) in \cite{Hergt:2008jn} (notice that eq.~(2.13) 
in \cite{Hergt:2008jn} has a typo), we obtain
\begin{align}\label{hsc}
H_{\text{SSS(BHB)}}^{\text{LO}}&=\frac{G}{m_1^2r^4}\left[\frac{3}{2}\left(\vec{S}_{1}^2\, 
\vec{S}_{2}\cdot\vec{n}\times\vec{p}_{1}
+\vec{S}_{1}\cdot\vec{n}\,\vec{S}_{2}\cdot\vec{S}_{1}\times\vec{p}_{1}
-5\left(\vec{S}_{1}\cdot\vec{n}\right)^2\vec{S}_{2}\cdot\vec{n}\times\vec{p}_{1}\right.\right.\nn\\
&\qquad\qquad\quad +\vec{n}\cdot\vec{S}_{1}\times\vec{S}_{2}\left(\vec{S}_{1}\cdot\vec{p}_{1}
-5\vec{S}_{1}\cdot\vec{n}\,\vec{p}_{1}\cdot\vec{n}\right)\nn\\
&\qquad\quad -\frac{3m_1}{2m_2}\left(\vec{S}_{1}^2\,
\vec{S}_{2}\cdot\vec{n}\times\vec{p}_2+2\vec{S}_{1}\cdot\vec{n}\,
\vec{S}_{2}\cdot\vec{S}_{1}\times\vec{p}_{2}
-5\left(\vec{S}_{1}\cdot\vec{n}\right)^2\vec{S}_{2}\cdot\vec{n}\times\vec{p}_{2}\right)\bigg)\nn\\
&\qquad\qquad \left.-\vec{S}_{1}\times\vec{n}\cdot\left(\vec{p}_{2}
-\frac{m_2}{4m_1}\vec{p}_{1}\right)\left(\vec{S}_{1}^2
-5\left(\vec{S}_{1}\cdot\vec{n}\right)^2\right)\right]\nn\\
&\quad+ 1\longleftrightarrow2.
\end{align}
We recall that we already fixed the gauge of the rotational variables to the canonical one.
Thus we note that the Legendre transform of the effective potentials at LO is trivial. 
The velocities are expressed in terms of the momenta, which at the LO level amounts to just 
using the Newtonian relation, i.e.~$v=p/m$. Also we can set in 
eq.~\eqref{vsc} the Wilson coefficients $C_{\text{ES$^2$}}=C_{\text{BS$^3$}}=1$
for the black hole case. 

Then we find for the difference between the LO cubic in spin potentials for 
the BHB case, 
$\Delta V_{\text{SSS(BHB)}}\equiv V_{\text{SSS(BHB)}}^{\text{ADM}}
-V_{\text{SSS(BHB)}}^{\text{EFT}}$,
which originates in our potential from diagrams 2(a), 2(b), and the extra 
addition, that it vanishes by virtue of the following vector identity,
valid for 4 arbitrary vectors $\vec{A}_a$ (with label index $a$) in 3 dimensions:
\begin{equation}\label{NidentityR}
N_i \equiv \frac{1}{3!} \epsilon_{abcd} \,\epsilon^{jkl} A^a_i A^b_j A^c_k A^d_l \equiv 0,
\end{equation}
namely $\vec{N}$ is a null vector \footnote{We thank an anonymous referee for providing eq.~\eqref{NidentityR} for the vector identity in eq.~\eqref{Nidentity}.}. Explicitly this reads
\begin{align}\label{Nidentity}
\vec{N}[\vec{A}_a] \equiv
\vec{A}_1 \;\, \vec{A}_2 \cdot \vec{A}_3 \times \vec{A}_4 
- \vec{A}_2 \;\, \vec{A}_3 \cdot \vec{A}_4 \times \vec{A}_1 
+ \vec{A}_3 \;\,\vec{A}_4 \cdot \vec{A}_1 \times \vec{A}_2 
- \vec{A}_4 \;\,\vec{A}_1 \cdot \vec{A}_2 \times \vec{A}_3
\equiv \vec{0}.
\end{align}
More specifically, we have for the difference 
\begin{align}
\Delta V_{\text{SSS(BHB)}}
=& -\frac{3}{2}\frac{G}{m_1r^4}\left(\vec{S}_1-5\vec{S}_1\cdot\vec{n}\,\vec{n}\right)\cdot 
\vec{N}[\vec{n},\vec{v}_1,\vec{S}_1,\vec{S}_2]
+ 1 \leftrightarrow 2 \equiv 0.
\end{align}
Therefore, our potential in eq.~\eqref{vsc} agrees with the result in eq.~\eqref{hsc} 
from \cite{Hergt:2007ha,Hergt:2008jn} 
for the case of binary black holes. One can also use this comparison 
to conclude that indeed $C_{\text{BS$^3$}} = 1$ for black holes.

\section{Leading order quartic in spin interaction} \label{loqsi}

\subsection{Feynman diagrams}

The quartic in spin interaction contains three kinds of interaction: 
a quadrupole-quadrupole interaction, an octupole-dipole one, 
and a hexadecapole-monopole one. The octupole-dipole interaction 
is analogous to the LO spin1-spin2, which is a dipole-dipole interaction, 
and as we noted the octupole moment corresponds to the dipole due to its 
odd parity. Then, the quadrupole-quadrupole and hexadecapole-monopole 
interactions are analogous each to the LO spin-squared interaction, which 
is a quadrupole-monopole interaction as we noted in section \ref{fsseft}, 
since the quadrupole and hexadecapole moments correspond to the monopole 
due to their even parity. We recall from figure 1 in \cite{Levi:2008nh}, 
that the LO spin1-spin2 interaction contains a single Feynman diagram, 
mediated by a one-graviton exchange of the gravito-magnetic vector. 
Moreover, we saw in figure 1 in section \ref{fsseft} here, that the LO 
spin-squared interaction also contains a single Feynman diagram mediated 
by a one-graviton exchange of the Newtonian scalar. Therefore, all in all 
we expect to have here three contributing Feynman diagrams, one for each 
of the three kinds of interaction, that make up the quartic in spin interaction. 

Indeed, the three contributing Feynman diagrams are shown here in figure 3, 
where on the left and right diagrams, (a) and (c), we have the quadrupole-quadrupole 
and hexadecapole-monopole interactions, and on the middle diagram, (b), 
we have the octupole-dipole interaction. 
These diagrams are obtained by the following contractions: 
in figure 3(a) the LO worldline spin quadrupole coupling to the Newtonian scalar 
from eq.~\eqref{eq:sqphi} is contracted with itself; in figure 3(b) we contract 
the LO worldline spin octupole coupling to the gravito-magnetic vector from 
eq.~\eqref{eq:s3A} with the LO spin coupling in eq.~\eqref{eq:sA}; finally, in 
figure 3(c) the LO worldline spin hexadecapole coupling to the Newtonian scalar 
from eq.~\eqref{eq:s4phi} is contracted with the LO mass coupling in eq.~\eqref{eq:mphi}. 

\begin{figure}[t]
\begin{center}
\includegraphics[scale=0.7]{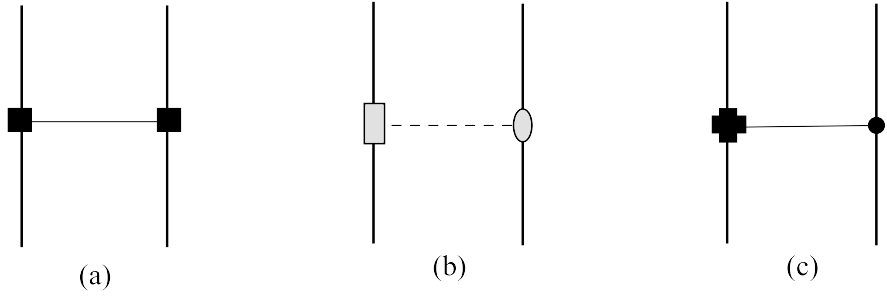}
\caption{LO quartic in spin interaction Feynman diagrams. 
Diagrams b and c here should be included together with their mirror images.
On the left and right we have the quadrupole-quadrupole and 
hexadecapole-monopole interactions, each of which is analogous 
to the LO spin-squared interaction in figure 1 here. On the 
middle we have the octupole-dipole interaction analogous to  
the LO spin1-spin2 interaction in figure 1 of \cite{Levi:2008nh}.} 
\end{center}
\label{fig:S4LO}
\end{figure}

Hence, the values of the Feynman diagrams of the LO 
quartic in spin interaction are given by the following:
\begin{align}
\text{Fig.~3(a)}=&\frac{3}{4}C_{1\text{(ES$^2$)}}C_{2(\text{ES$^2$)}}
\frac{G}{m_1m_2r^5}\left[S_1^2S_2^2+2\left(\vec{S}_1\cdot\vec{S}_2\right)^2
-5\left(S_1^2\left(\vec{S}_2\cdot\vec{n}\right)^2\right.\right.\nn\\
&\qquad\qquad\qquad \left.\left.+S_2^2\left(\vec{S}_1\cdot\vec{n}\right)^2
+4\vec{S}_1\cdot\vec{S}_2\,\vec{S}_1\cdot\vec{n}\,\vec{S}_2\cdot\vec{n}
-7\left(\vec{S}_1\cdot\vec{n}\right)^2\left(\vec{S}_2\cdot\vec{n}\right)^2\right)\right],\\
\text{Fig.~3(b)}=&\frac{3}{2}C_{1\text{(BS$^3$)}}\frac{G}{m_1^2r^5}
\left[S_1^2\,\vec{S}_1\cdot\vec{S}_2-5S_1^2\,\vec{S}_1\cdot\vec{n}\,\vec{S}_2\cdot\vec{n}
-5\vec{S}_1\cdot\vec{S}_2\left(\vec{S}_1\cdot\vec{n}\right)^2\right.\nn\\
&\qquad\qquad\qquad\qquad\left.+\frac{35}{3}\vec{S}_2\cdot\vec{n}
\left(\vec{S}_1\cdot\vec{n}\right)^3\right],\\
\text{Fig.~3(c)}=&\frac{3}{8}C_{1\text{(ES$^4$)}}\frac{Gm_2}{m_1^3r^5}
\left[S_1^4-10S_1^2\left(\vec{S}_1\cdot\vec{n}\right)^2
+\frac{35}{3}\left(\vec{S}_1\cdot\vec{n}\right)^4\right].
\end{align}
Here too the evaluation of the diagrams is straightforward. 
Moreover, we have no additions to the effective potential.

\subsection{Effective potential and Hamiltonian}

Summing all diagrams in figure 3, we get the following 
effective potential for the complete LO quartic in spin interaction:
\begin{align} \label{vsq}
V_{\text{SSSS}}^{\text{LO}}=&-\frac{3}{4}C_{1\text{(ES$^2$)}}
C_{2\text{(ES$^2$)}}\frac{G}{m_1m_2r^5}\left[S_1^2S_2^2
+2\left(\vec{S}_1\cdot\vec{S}_2\right)^2
-5\left(S_1^2\left(\vec{S}_2\cdot\vec{n}\right)^2\right.\right.\nn\\
&\qquad\qquad\qquad\quad \left.\left.+S_2^2\left(\vec{S}_1\cdot\vec{n}\right)^2
+4\vec{S}_1\cdot\vec{S}_2\,\vec{S}_1\cdot\vec{n}\,\vec{S}_2\cdot\vec{n}
-7\left(\vec{S}_1\cdot\vec{n}\right)^2\left(\vec{S}_2\cdot\vec{n}\right)^2\right)\right]\nn\\
&-\frac{3}{2}C_{1\text{(BS$^3$)}}\frac{G}{m_1^2r^5}\left[S_1^2\,\vec{S}_1\cdot\vec{S}_2
-5S_1^2\,\vec{S}_1\cdot\vec{n}\,\vec{S}_2\cdot\vec{n}
-5\vec{S}_1\cdot\vec{S}_2\left(\vec{S}_1\cdot\vec{n}\right)^2\right.\nn\\
&\qquad\qquad\qquad\qquad\left.+\frac{35}{3}\vec{S}_2\cdot\vec{n}
\left(\vec{S}_1\cdot\vec{n}\right)^3\right]+1\longleftrightarrow2\nn\\
&-\frac{3}{8}C_{1\text{(ES$^4$)}}\frac{Gm_2}{m_1^3r^5}\left[S_1^4
-10S_1^2\left(\vec{S}_1\cdot\vec{n}\right)^2
+\frac{35}{3}\left(\vec{S}_1\cdot\vec{n}\right)^4\right]+1\longleftrightarrow2.
\end{align}
Here too, for black holes the new Wilson coefficient $C_{\text{ES}^4}$ 
can be fixed from the gauge invariant binding energy in 
\cite{Steinhoff:2012rw}, which also checks our result in the 
small mass ratio case. From this we find that $C_{\text{ES}^4}=1$ 
for black holes.

We proceed to compare our effective potential with the ADM Hamiltonian 
results for a black hole binary in \cite{Hergt:2007ha,Hergt:2008jn}.
However, it was found in \cite{Steinhoff:2012rw}, that the black hole 
binary Hamiltonian at quartic order in each of the spins, which was 
derived in \cite{Hergt:2008jn}, must be incomplete. Indeed, at leading order 
the source part of the Hamilton constraint $\mathcal{H}^{\text{matter}}$ 
is the source of the Newtonian potential, which corresponds to the NRG scalar 
field $\phi$. From eq.~\eqref{eq:s4phi} we therefore expect a contribution 
to the Hamilton constraint of the form:
\begin{equation} \label{corhadm}
\mathcal{H}^{\text{matter}}_{\text{hexadecapole}}
 = \frac{C_{ES^4}}{24m^3}S^{i}S^{j}S^{k}S^{l}
 \partial_i\partial_j\partial_k\partial_l\delta .
\end{equation}
Indeed, this term was not considered in \cite{Hergt:2008jn}.
The resulting contribution to the Hamiltonian is identical 
to the value of figure 3(c) up to an overall sign.
Hence, we see that the conclusions of section VI and in particular 
eq.~(6.5) in \cite{Hergt:2008jn} are incorrect. 

Collecting the pieces from eqs.~(124), (131) in \cite{Hergt:2007ha}, 
and taking into account our correction to eq.~(6.5) in \cite{Hergt:2008jn}, 
that we just noted in eq.~\eqref{corhadm}, coming from the hexadecapole-monopole interaction in 
fig.~3(c) here (for $C_{\text{ES}^4} = 1$), we obtain the complete correct binary 
black hole ADM Hamiltonian:
\begin{align} \label{corhsq}
H_{\text{SSSS(BHB)}}^{\text{LO}}&=-\frac{3}{2}\frac{G}{m_{1}m_{2}r^5}
\left[\frac{1}{2}\vec{S}_{1}^2\vec{S}_{2}^2+\left(\vec{S}_{1}\!\cdot\!\vec{S}_{2}\right)^2
-\frac{5}{2}\left(\vec{S}_{1}^2\left(\vec{S}_{2}\!\cdot\!\vec{n}\right)^2
+\vec{S}_{2}^2\left(\vec{S}_{1}\!\cdot\!\vec{n}\right)^2\right)\right.\nn\\
&\left.\qquad\qquad\qquad -10\vec{S}_{1}\!\cdot\!\vec{n}\,\vec{S}_{2}\!\cdot\!\vec{n}
\left(\vec{S}_{1}\!\cdot\!\vec{S}_{2}
-\frac{7}{4}\vec{S}_{1}\!\cdot\!\vec{n}\,\vec{S}_{2}\!\cdot\!\vec{n}\right)\right]\nn\\
&\quad-\frac{3}{2}\frac{G}{m_{1}^2r^5}\left[\vec{S}_{1}^2\,\vec{S}_{1}\!\cdot\!\vec{S}_{2}
-5\vec{S}_{1}\!\cdot\!\vec{S}_{2}\left(\vec{S}_{1}\!\cdot\!\vec{n}\right)^2
-5\vec{S}_{1}^2\,\vec{S}_{1}\!\cdot\!\vec{n}\,\vec{S}_{2}\!\cdot\!\vec{n}\right.\nn\\
&\qquad\qquad\qquad\qquad\quad +\frac{35}{3}\vec{S}_{2}\!\cdot\!\vec{n}\left(\vec{S}_{1}
\!\cdot\!\vec{n}\right)^3\bigg]
+1\longleftrightarrow2\nn\\
&\quad-\frac{3Gm_2}{8m_1^3r^5}\left[S_1^4-10S_1^2\left(\vec{S}_1\cdot\vec{n}\right)^2
+\frac{35}{3}\left(\vec{S}_1\cdot\vec{n}\right)^4\right]+1\longleftrightarrow2.
\end{align}
With our correction included we then find full agreement of 
our result in eq.~\eqref{vsq} with the black hole binary ADM Hamiltonian in 
eq.~\eqref{corhsq}. Again, this comparison can 
also be used to fix $C_{\text{BS}^3}=1$ for the black hole case.

\section{Conclusions} \label{theendmyfriend}

In this work we derived for the first time the complete LO cubic and 
quartic in spin interaction potentials for generic compact binaries 
via the effective field theory for gravitating spinning objects \cite{Levi:2015msa}. 
These corrections, 
which enter at the 3.5PN and 4PN orders, respectively, for rapidly rotating 
compact objects, complete the LO finite size effects with spin up to the 
4PN accuracy. In order to complete the spin dependent conservative sector to 4PN order, it remains to apply the effective field theory for gravitating spinning objects \cite{Levi:2015msa} at NNLO to quadratic level in spin, which was initiated in \cite{Levi:2011eq}.

We arrive at the results here by augmenting the effective action with 
new higher dimensional nonminimal coupling worldline operators, 
involving higher-order derivatives of the field, coupled to the 
higher-order multipole moments with spins, and introducing new Wilson coefficients, 
corresponding to constants, which describe the octupole and hexadecapole 
deformations of the object due to spin. These Wilson coefficients are fixed 
to unity in the black hole case via comparisons with the gauge invariant binding 
energy in the test particle limit and with the ADM Hamiltonian. We also see that the 
ADM Hamiltonian result for the quartic in spin interaction potential for a black hole binary, 
which was derived in \cite{Hergt:2008jn}, is incorrect, and we complete this result.  

It should be noted that the relation between the Wilson coefficients in this 
work and the multipole moments used in numerical codes, 
e.g.~\cite{Laarakkers:1997hb, Pappas:2012ns, Yagi:2014bxa}, and the Geroch-Hansen multipoles for black holes noted in eq.~\eqref{GHansen}, should be worked
out via a formal EFT matching procedure. Clearly, this involves subtleties 
\cite{Pappas:2012ns}, and is left for future research.
Yet we also note that recently, universal relations, nearly equation-of-state independent, 
were found between certain neutron star observables \cite{Yagi:2013bca, Yagi:2013awa}.
It was demonstrated in \cite{Yagi:2014bxa}, that all multipoles seem to be 
related in an approximately universal manner, though the accuracy of this 
approximation becomes worse for higher multipoles.
This implies a universal relation between the coefficients $C_{ES^2}$, $C_{BS^3}$,
and $C_{ES^4}$ in our potentials also for neutron stars. 
Thus, the new potentials effectively do not introduce new parameters. This is 
important for gravitational wave experiments, since a larger parameter space 
would render the parameter estimation more difficult. Instead, our potentials
essentially just refine the precision of the equations of motion or of the 
binding energy without enlarging the parameter space.
 
The nonminimal coupling worldline operators enter the point particle action 
with the electric and magnetic components of the Weyl tensor of even and odd parity, 
coupled to the even and odd worldline spin tensors, respectively.
Moreover, the NRG field decomposition, which we employ, demonstrates 
a coupling hierarchy of the gravito-magnetic vector and the Newtonian scalar 
to the odd and even in spin operators, respectively, which extends 
that of minimal coupling. Therefore the NRG fields are found 
to be very useful for the treatment of interactions involving spins, 
since they also facilitate the construction of the Feynman diagrams, 
and provide here instructive analogies between the LO spin-orbit 
and cubic in spin interactions, and between the LO 
quadratic and quartic in spin interactions. These analogies are based on 
the correspondence between the even and odd parity multipole moments with spin. 

Finally, we note that we see that beyond the LO finite size effect 
with spin, namely the LO spin-squared interaction, which is a purely 
Newtonian effect, all LO finite size effects with spin are relativistic ones.

\acknowledgments

This work has been done within the Labex ILP (reference ANR-10-LABX-63) part 
of the Idex SUPER, and received financial French state aid managed by the 
Agence Nationale de la Recherche, as part of the programme Investissements 
d'Avenir under the reference ANR-11-IDEX-0004-02.

\paragraph{Note added.} 
After the preprint of this work has appeared, there appeared two related works: 
One work \cite{Vaidya:2014kza} recovered specific pieces of the LO cubic and quartic in spin 
Hamiltonians from the S-matrix combined with EFT techniques;
Another work \cite{Marsat:2014xea} computed LO cubic in spin effects at the level of the EOM, and at the 
energy flux, via traditional methods. 
Both works found agreement with our results.

\bibliographystyle{jhep}
\bibliography{gwbibtex}

\end{document}